\documentclass[11pt]{article}
\usepackage{amssymb}

\usepackage[totalwidth=460truept,totalheight=600truept]{geometry}
\usepackage{latexsym,amssymb,amsmath,graphicx,accents,eucal,slashed,subfigure}
\usepackage{dsfont}
\usepackage{hyperref}

\hypersetup{hypertexnames=false}
\def\theequation{\arabic{section}.\arabic{equation}}

\global\arraycolsep=1truept
\numberwithin{equation}{section}
\renewcommand{\theequation}{\arabic{section}.\arabic{equation}}

\begin{document}

\bigskip \hfill IFUP-TH 2013/07

\vskip 1.4truecm

\begin{center}
{\huge \textbf{\vspace{10pt}Properties Of The Classical Action}}\vskip %
.5truecm {\huge \textbf{\vspace{10pt}Of Quantum Gravity}} \vskip 1truecm

\textsl{Damiano Anselmi}

\vskip .2truecm

\textit{Dipartimento di Fisica ``Enrico Fermi'', Universit\`{a} di Pisa, }

\textit{and INFN, Sezione di Pisa,}

\textit{Largo B. Pontecorvo 3, I-56127 Pisa, Italy}

\vskip .2truecm

damiano.anselmi@df.unipi.it

\vskip 1.5truecm

\textbf{Abstract}
\end{center}

\medskip

{\small The classical action of quantum gravity, determined by
renormalization, contains infinitely many independent couplings and can be
expressed in different perturbatively equivalent ways. We organize it in a
convenient form, which is based on invariants constructed with the Weyl
tensor. We show that the FLRW metrics are exact solutions of the field
equations in arbitrary dimensions, and so are all locally conformally flat
solutions of the Einstein equations. Moreover, expanding the metric tensor
around locally conformally flat backgrounds the quadratic part of the action
is free of higher derivatives. Black-hole solutions of Schwarzschild and
Kerr type are modified in a non-trivial way. We work out the first
corrections to their metrics and study their properties. }

\vspace{4pt}

\vskip 1truecm

\vfill\eject

\section{Introduction}

\setcounter{equation}{0}

Having control on the classical action of quantum gravity, its properties
and the solutions of its field equations can be useful to address the search
for detectable effects that may single out some significant departure from
Einstein gravity. Since quantum gravity is not power-counting
renormalizable, its classical action contains infinitely many independent
couplings. Nevertheless, some interesting solutions of the field equations
may depend only on a finite subset of parameters and allow us to make
physical predictions. Moreover, even if each correction is small, the
presence of infinitely many of them opens the door to effects that might be
detectable in particular experimental arrangements or astrophysical
observations, situated beyond the domains tested so far and before radiative
corrections become important. Finally, in extreme situations, such as inside
black holes, or close to the event horizon, or in the primordial phases of
the universe, classical corrections of quantum origin may play a relevant
role.

Using field redefinitions, the classical action of quantum gravity can be
written in different, perturbatively equivalent expansions around the
Einstein action. In particular, we can rearrange the terms proportional to
the Einstein vacuum field equations. Equivalent actions can be useful to
uncover different classes of exact solutions of the field equations, or
reduce the effort to study approximate solutions.

In this paper we single out a form $S_{\text{QG}}$ that we deem convenient
for several purposes. Besides the Hilbert term, the cosmological term and a
peculiar combination that is non-trivial in higher dimensions, the action $%
S_{\text{QG}}$ contains invariants constructed with the Weyl tensor $C_{\mu
\nu \rho \sigma }$, rather than the Riemann tensor $R_{\mu \nu \rho \sigma }$%
. Precisely, we write, in arbitrary dimensions $d>2$, 
\begin{equation}
S_{\text{QG}}=-\frac{1}{2\kappa ^{d-2}}\int \sqrt{-g}\left( R+2\Lambda
+\lambda _{0}\kappa ^{2}\mathrm{\hat{G}}+\lambda _{1}\kappa ^{4}C_{_{{3}%
}}+\lambda _{1}^{\prime }\kappa ^{4}C_{3}^{\prime }+\sum_{n=2}^{\infty
}\lambda _{n}\kappa ^{2n+2}\mathcal{I}_{n}(\nabla ,C)\right) +S_{m},
\label{cqgac}
\end{equation}
where $\kappa $ has dimension $-1$ in units of mass, $\lambda _{n}$ are
dimensionless constants, $S_{m}$ are the contributions of matter fields and
other gauge fields, $\mathrm{\hat{G}}$ is the special combination \cite{abse}
\[
\mathrm{\hat{G}}=R_{\mu \nu \rho \sigma }R^{\mu \nu \rho \sigma }-4R_{\mu
\nu }R^{\mu \nu }+R^{2}+\frac{4(d-3)(d-4)}{(d-1)(d-2)}\Lambda \left(
R+\Lambda \right) , 
\]
and $\mathcal{I}_{n}(\nabla ,C)$ collectively denotes the local scalars of
dimension $2n+4$ that can be constructed with three or more Weyl tensors $%
C_{\mu \nu \rho \sigma }$ and covariant derivatives $\nabla _{\mu }$, up to
covariant divergences of vectors. Each such scalar must be multiplied by an
independent $\lambda _{n}$. For future use, we explicitly write the terms $%
\mathcal{I}_{1}(\nabla ,C)$, which are two contractions of three Weyl
tensors: 
\[
C_{3}=C_{\mu \nu \rho \sigma }C^{\rho \sigma \alpha \beta }C_{\alpha \beta
}^{\hspace{0.12in}\mu \nu },\qquad C_{3}^{\prime }=C_{\mu \rho \nu \sigma
}C^{\alpha \mu \beta \nu }C_{\hspace{0.06in}\alpha \hspace{0.07in}\beta
}^{\rho \hspace{0.07in}\sigma }. 
\]
For simplicity in this paper we assume parity invariance. Parity-violating
terms may be treated along the same guidelines.

In four dimensions $\sqrt{-g}\mathrm{\hat{G}}$ is the Gauss-Bonnet
integrand, which does not contribute to the field equations. Thanks to this
fact, quantum gravity is finite at one loop in the absence of matter \cite
{thooftveltman}. Moreover, $C_{3}$ and $C_{3}^{\prime }$ are proportional to
each other, so we can set $\lambda _{1}^{\prime }=0$ in $d=4$. Goroff and
Sagnotti showed \cite{sagnotti} that $C_{3}$ is switched on as a two-loop
counterterm in pure gravity. The result of their calculation can be
interpreted as the running of the coupling constant $\lambda _{1}$,
therefore allows us to infer that quantum gravity predicts $\lambda _{1}\neq
0$. In principle, the presence of matter can modify this conclusion, but
only if matter fields exactly cancel the $C_{3}$-counterterm generated by
pure gravity, and the cancellation is consistent with renormalization-group
invariance. As far as we know today, this happens only in supergravity.
Similarly, $C_{3}$ is turned on at one loop in six dimensional pure gravity 
\cite{vannieu}.

The action (\ref{cqgac}) is preserved by renormalization. It is
perturbatively equivalent to actions written previously and to the most
general local perturbative extension of the Einstein action \cite{wein}. The
form (\ref{cqgac}) is convenient in various respects, for example it allows
us to find interesting classes of exact solutions of the field equations,
which include all locally conformally flat metrics (which we just call
``conformally flat'' from now on) that solve the Einstein equations. In
particular, the Friedmann-Lema\^{i}tre-Robertson-Walker (FLRW) metrics are
exact solutions of the $S_{\text{QG}}$-field equations in arbitrary
dimensions $d>2$ with a homogeneous and isotropic matter distribution. In
four dimensions such solutions coincide with the usual ones, while in higher
dimensions they coincide with the usual solutions once the energy density $%
\rho $ and the pressure $p$ are replaced by suitable functions of $\rho $
and $p$. Metric-independent maps also relate conformally flat solutions of
the $S_{\text{QG}}$-equations to conformally flat solutions of the Einstein
equations. On the other hand, solutions that are not conformally flat are
deformed in a nontrivial way by the couplings $\lambda _{n}$. In the paper
we study the first modifications to the Schwarzschild and Kerr metrics in
four dimensions.

Another property of $S_{\text{QG}}$ is that expanding the metric tensor
around conformally flat backgrounds the quadratic part of the action is free
of higher derivatives. Vertices, instead, as well as quadratic terms
obtained expanding around more general backgrounds, do not have this
feature. Working perturbatively in the couplings $\lambda _{n}$, every term
of the field equations that contains higher derivatives can be converted into a linear combination
of terms that contain at most two derivatives. Then the solutions of the $S_{%
\text{QG}}$-field equations are uniquely determined by their $\lambda
_{n}\rightarrow 0$-limits.

The paper is organized as follows. In section 2 we study the action (\ref
{cqgac}) and its field equations. We compare $S_{\text{QG}}$ with other
local perturbative extensions of the Einstein action. In section 3 we study
exact solutions of the $S_{\text{QG}}$-field equations in arbitrary
dimensions, in particular metrics of FLRW type and conformally flat metrics.
In section 4 we work out the first corrections to the Schwarzschild and Kerr
black-hole solutions in four dimensions and discuss their properties. In
section 5 we study the perturbative equivalence of actions in detail.
Section 6 collects our conclusions. In the appendix we show how to truncate
the actions to finite numbers of terms, consistently with the diagrammatic
expansion of quantum gravity.

\section{The action and its field equations}

\setcounter{equation}{0}

In this section we study $S_{\text{QG}}$ and compare its properties with
those of two other actions: the most general local perturbative extension $%
S_{\text{loc}}$ of the Einstein action \cite{wein} and an action written in
ref. \cite{abse}, which inspires the simplification proposed here. It is
convenient to parametrize $S_{\text{loc}}$ as 
\begin{equation}
S_{\text{loc}}=-\frac{1}{2\kappa ^{d-2}}\int \sqrt{-g}\left( R+2\Lambda
+\sum_{n=0}^{\infty }\bar{\lambda}_{n}\kappa ^{2n+2}\mathcal{\bar{I}}%
_{n}^{(\Lambda )}(\nabla ,\hat{R})\right) +S_{m},  \label{sloc}
\end{equation}
where $\mathcal{\bar{I}}_{n}^{(\Lambda )}(\nabla ,\hat{R})$ denotes the
scalars of dimensions $2n+4$ that can be constructed with two or more
tensors 
\begin{equation}
\hat{R}_{\mu \nu \rho \sigma }=R_{\mu \nu \rho \sigma }+\frac{2\Lambda }{%
\left( d-1\right) \left( d-2\right) }\left( g_{\mu \rho }g_{\nu \sigma
}-g_{\mu \sigma }g_{\nu \rho }\right) ,  \label{hatc}
\end{equation}
as well as their contractions $\hat{R}_{\mu \nu }=\hat{R}_{\hspace{0.07in}%
\mu \rho \nu }^{\rho }$ and $\hat{R}=\hat{R}_{\mu }^{\mu }$, and covariant
derivatives $\nabla _{\mu }$, up to covariant divergences of vectors.

In ref. \cite{abse} the properties of renormalization were used to write a
different action, namely\footnote{%
Up to notational changes.}

\begin{equation}
\tilde{S}_{\text{QG}}=-\frac{1}{2\kappa ^{d-2}}\int \sqrt{-g}\left(
R+2\Lambda +\tilde{\lambda}_{0}\kappa ^{2}\mathrm{\hat{G}}%
+\sum_{n=1}^{\infty }\tilde{\lambda}_{n}\kappa ^{2n+2}\mathcal{I}%
_{n}^{(\Lambda )}(\nabla ,\hat{R})\right) +S_{m},  \label{oldac}
\end{equation}
where $\mathcal{I}_{n}^{(\Lambda )}(\nabla ,\hat{R})$ denotes the local
scalars of dimensions $2n+4$ that can be constructed with three or more
tensors $\hat{R}_{\mu \nu \rho \sigma }$ and covariant derivatives $\nabla
_{\mu }$, up to covariant divergences of vectors. The contractions $\hat{R}%
_{\mu \nu }$ and $\hat{R}$ can appear inside the scalars $\mathcal{I}%
_{n}^{(\Lambda )}$ or not, the resulting different actions being
perturbatively equivalent (see section 5 for more details).

The actions $S_{\text{QG}}$ and $\tilde{S}_{\text{QG}}$ look like
restrictions on $S_{\text{loc}}$, but they are actually perturbatively
equivalent to each other and to $S_{\text{loc}}$. Precisely, these actions
can be mapped into one another by means of local field redefinitions and
parameter-redefinitions, the parameters $\lambda _{n}$, $\tilde{\lambda}_{n}$
and $\bar{\lambda}_{n}$ being treated perturbatively. Consequently, $S_{%
\text{QG}}$ and $\tilde{S}_{\text{QG}}$ are preserved by renormalization,
namely all divergences generated by Feynman diagrams can be subtracted
redefining the metric tensor and the parameters $\lambda _{n}$, or $\tilde{%
\lambda}_{n}$, the matter fields and the parameters contained inside $S_{m}$.

Specifically, the renormalizability of $S_{\text{loc}}$ is
obvious, since it is the most general local action. Instead, the actions $S_{\text{QG}}$ and $\tilde{S}_{\text{QG}}$ are renormalizable,
since using Bianchi identities, commuting covariant derivatives and
integrating by parts every (counter)term that does not appear in those
actions can be reabsorbed away redefining fields and parameters \cite{abse}.
In particular, the scalar $\mathrm{\hat{G}}$ is used to write $\hat{R}_{\mu
\nu \rho \sigma }\hat{R}^{\mu \nu \rho \sigma }$ as a linear combination of
terms present in $S_{\text{QG}}$ ($\tilde{S}_{\text{QG}}$), plus terms
quadratically proportional to $\hat{R}_{\mu \nu }$. In turn, these can be
converted into terms of $S_{\text{QG}}$ ($\tilde{S}_{\text{QG}}$) redefining the metric tensor. 

Spaces of constant curvature play a peculiar role, since in the absence of
matter they are exact solutions of the field equations of the most general
action. Indeed, once $R_{\mu \nu \rho \sigma }=K(g_{\mu \rho }g_{\nu \sigma
}-g_{\mu \sigma }g_{\nu \rho })$ is used, with $K=$constant, any covariant
gravitational field equations must reduce to a simple condition 
\begin{equation}
f(\kappa ^{2}K,\kappa ^{2}\Lambda ,\lambda )g_{\mu \nu }=0,  \label{doingso}
\end{equation}
where $f$ is some function of the parameters of the theory, which can be
solved to obtain $K$. The parametrizations of (\ref{cqgac}), (\ref{sloc})
and (\ref{oldac}), which use hatted tensors or Weyl tensors, are such that
the solution of (\ref{doingso}) simply reads 
\begin{equation}
K=-\frac{2\Lambda }{\left( d-1\right) \left( d-2\right) }.  \label{constcurv}
\end{equation}

An important fact is that $S_{\text{QG}}$ and $\tilde{S}_{\text{QG}}$,
differently from $S_{\text{loc}}$, do not contain terms that are quadratic
in the curvature tensors, with the exception of those appearing in the
peculiar combination $\mathrm{\hat{G}}$. The special scalar $\mathrm{\hat{G}}
$ is a generalization of the Gauss-Bonnet integrand. Its main property is
that expanding the metric around a background $\bar{g}_{\mu \nu }$ of
constant curvature $K$ equal to (\ref{constcurv}), the integral $\int \sqrt{%
-g}\mathrm{\hat{G}}$ does not contain terms that are linear or quadratic in
the quantum fluctuations. Precisely, writing $g_{\mu \nu }=\bar{g}_{\mu \nu
}+h_{\mu \nu }$ and using $\hat{R}_{\mu \nu \rho \sigma }(\bar{g})=0$ it is
straightforward to check that 
\begin{equation}
\int \sqrt{-g}\mathrm{\hat{G}}=\frac{32(d-3)\Lambda ^{2}}{(d-1)(d-2)^{2}}%
\int \sqrt{-\bar{g}}+\mathcal{O}\left( h^{3}\right) .  \label{bibi}
\end{equation}
The invariants $\int \sqrt{-g}\mathcal{I}_{n}$ and $\int \sqrt{-g}\mathcal{I}%
_{n}^{(\Lambda )}$, $n\geqslant 1$, clearly have the same property. Thus, in
this expansion the quadratic parts of the actions $S_{\text{QG}}$ and $%
\tilde{S}_{\text{QG}}$ do not contain higher derivatives and coincide with
the quadratic part obtained from Einstein gravity. The absence of higher
derivatives in propagators is important to prevent the propagation of
unphysical degrees of freedom, such as those of higher-derivative quantum
gravity \cite{stelle}.

In every even dimensions $d=2k$ we can drop one term $\sim \int \sqrt{-g}%
C^{k}$ containing $k$ Weyl tensors and no derivatives and add the
topological invariant 
\begin{equation}
\int \sqrt{-g}\mathrm{G}_{k}\equiv \int \sqrt{-g}\hspace{0.01in}\delta _{\mu
_{1}\nu _{1}\cdots \mu _{k}\nu _{k}}^{\alpha _{1}\beta _{1}\cdots \alpha
_{k}\beta _{k}}R_{\hspace*{0.1in}\hspace*{0.1in}\alpha _{1}\beta _{1}}^{\mu
_{1}\nu _{1}}\hfill \cdots R_{\hspace*{0.1in}\hspace*{0.1in}\alpha _{k}\beta
_{k}}^{\mu _{k}\nu _{k}}  \label{lovelock}
\end{equation}
instead, which does not contribute to the field equations. The difference
between two such actions is a linear combination of other terms $\sim \int 
\sqrt{-g}C^{k}$ plus terms containing the Ricci tensor \cite{vannieu2}.
Writing the Ricci tensor as a linear combination of $\hat{R}_{\mu \nu }$ and 
$\Lambda g_{\mu \nu }$, we can reabsorb the difference into a perturbative
local field redefinition and parameter-redefinitions (see section 5). For
example, in six dimensions we can set $\lambda _{1}^{\prime }=0$ and add $%
\int \sqrt{-g}\mathrm{G}_{3}$.

The invariants (\ref{lovelock}) with $k<d/2$ are not topological.
Nevertheless, their variations with respect to the metric tensor are free of
higher derivatives \cite{lovelock}. The action of Lovelock gravity \cite
{lovelock} in $d$ dimensions contains only the invariants (\ref{lovelock})
with $k\leqslant d/2$, therefore its field equations are completely free of
higher derivatives. Nevertheless, that kind of action is not preserved by
renormalization. For example, in four dimensions Lovelock gravity is just
Einstein gravity with the Gauss-Bonnet term.

In our approach, which is based on renormalization, the mentioned property
of $\mathrm{G}_{k}$ is used only for $k=2$ \cite{zwiebach,abse}, to build
the special invariant $\int \sqrt{-g}\mathrm{\hat{G}}$ and ensure that when
we expand the metric around backgrounds of special classes, the quadratic
parts of actions such as $S_{\text{QG}}$ and $\tilde{S}_{\text{QG}}$ are
free of higher derivatives. However, we cannot guarantee similar results for
vertices, or for the quadratic parts obtained expanding around more general
backgrounds.

The matter action $S_{m}$ is the most general local one, as long as it has
correct unitary propagators. If the classical action has correct
propagators, the renormalized one also has. Indeed, in a quantum field
theory of matter fields of spins $\leqslant 1/2$ and gauge fields of spins $%
\leqslant 2$, higher-derivative quadratic terms are not turned on by
renormalization if they are absent at the tree level \cite{abse}. This fact
ensures that a unitary propagator is not driven by renormalization into a
non-unitary one.

The new form $S_{\text{QG}}$ of the classical action improves $\tilde{S}_{%
\text{QG}}$ in various respects. First, the $\tilde{S}_{\text{QG}}$-scalars $%
\mathcal{I}_{n}^{(\Lambda )}$ are intrinsically $\Lambda $-dependent, being
constructed with hatted curvature tensors. This gives the impression that
the action $\tilde{S}_{\text{QG}}$ is chosen ad hoc. It is better to have
independent terms multiplied by independent couplings, as in $S_{\text{QG}}$%
. Moreover, (\ref{cqgac}) allows us to easily find other, more interesting
exact solutions of the field equations, besides spaces of constant
curvature, such as the FLRW metrics. More generally, all conformally flat
solutions of the Einstein equations solve the $S_{\text{QG}}$-field
equations (in four dimensions) or can be easily mapped into solutions of the 
$S_{\text{QG}}$-field equations (in higher dimensions). Finally, expanding
the metric tensor $g_{\mu \nu }$ around any conformally flat background $%
\bar{g}_{\mu \nu }$ the quadratic part of the action $S_{\text{QG}}$ is free
of higher derivatives. The action $\tilde{S}_{\text{QG}}$ satisfies this
property only for the expansion around spaces of constant curvature.

\paragraph{Field equations\newline
}

Writing 
\[
S_{\text{QG}}=-\frac{1}{2\kappa ^{d-2}}\int \sqrt{-g}\left( R+2\Lambda
\right) +S_{m}+S^{(g)}, 
\]
the $S_{\text{QG}}$-field equations read 
\begin{equation}
R_{\mu \nu }-\frac{1}{2}Rg_{\mu \nu }-\Lambda g_{\mu \nu }=\kappa
^{d-2}T_{\mu \nu }+\kappa ^{d-2}T_{\mu \nu }^{(g)},  \label{fieldequations}
\end{equation}
where 
\[
T_{\mu \nu }=\frac{2}{\sqrt{-g}}\frac{\delta S_{m}}{\delta g^{\mu \nu }}%
,\qquad T_{\mu \nu }^{(g)}=\frac{2}{\sqrt{-g}}\frac{\delta S^{(g)}}{\delta
g^{\mu \nu }}, 
\]
are the matter energy-momentum tensor and the gravitational
self-energy-momentum tensor, respectively. Varying $\int \sqrt{-g}\mathrm{%
\hat{G}}$ explicitly, we find 
\begin{eqnarray}
\kappa ^{d-2}T_{\mu \nu }^{(g)} &=&-\lambda _{0}\kappa ^{2}\left[ 2C_{\mu
\rho \sigma \alpha }C_{\nu }^{\hspace{0.07in}\rho \sigma \alpha }-\frac{1}{2}%
g_{\mu \nu }C_{\rho \sigma \alpha \beta }C^{\rho \sigma \alpha \beta }-\frac{%
4(d-4)}{d-2}C_{\mu \rho \nu \sigma }\hat{R}^{\rho \sigma }\right.  \nonumber
\\
&&\left. -\frac{(d-3)(d-4)}{(d-2)^{2}}\left( 4\hat{R}_{\mu \rho }\hat{R}%
_{\nu }^{\rho }-2g_{\mu \nu }\hat{R}_{\rho \sigma }\hat{R}^{\rho \sigma }-%
\frac{2d}{d-1}\hat{R}_{\mu \nu }\hat{R}+\frac{d+2}{2(d-1)}g_{\mu \nu }\hat{R}%
^{2}\right) \right]  \nonumber \\
&&+\mathcal{O}(\nabla ^{2}C^{2})+\mathcal{O}(RC^{2})+\mathcal{O}(C^{3}).
\label{tg}
\end{eqnarray}

The field equations of $\tilde{S}_{\text{QG}}$ are very similar, the only
difference being that in the third line of (\ref{tg}) the Weyl tensors are
replaced by hatted curvature tensors. The notation $\mathcal{O}(\hat{R}^{n})$
means terms containing at least $n$ powers of $\hat{R}_{\mu \nu \rho \sigma
} $ and its contractions.

Observe that the variation of $\int \sqrt{-g}\mathrm{\hat{G}}$ with respect
to the metric is $\mathcal{O}(\hat{R}^{2})$, in agreement with (\ref{bibi}).
Clearly, the tensor $T_{\mu \nu }^{(g)}$ of (\ref{tg}) is identically zero in three dimensions. In
four dimensions, instead, it reduces to the last line of (\ref{tg}). For
future use we explicitly work out the first non-trivial contributions to $%
T_{\mu \nu }^{(g)}$ in $d=4$, which are the ones proportional to the \textit{%
Goroff-Sagnotti constant} $\Lambda _{\text{GS}}\equiv 3\lambda _{1}\kappa
^{4}$. Setting $\lambda _{1}^{\prime }=0$ and dropping the Gauss-Bonnet
term, we write the four dimensional action as 
\[
S_{\text{QG}}^{(d=4)}=-\frac{1}{2\kappa ^{2}}\int \sqrt{-g}\left( R+2\Lambda
+\frac{\Lambda _{\text{GS}}}{3}C_{3}+\sum_{n=2}^{\infty }\lambda _{n}\kappa
^{2n+2}\mathcal{I}_{n}(\nabla ,C)\right) +S_{m}. 
\]
Then we find 
\begin{eqnarray}
\kappa ^{2}T_{\mu \nu }^{(g)} &=&\Lambda _{\text{GS}}\left( \nabla ^{\rho
}\nabla ^{\sigma }C_{\mu \rho \sigma \nu }^{(2)}+\nabla ^{\rho }\nabla
^{\sigma }C_{\nu \rho \sigma \mu }^{(2)}-\frac{1}{2}C_{\mu \alpha \rho
\sigma }^{(2)}R_{\nu }^{\hspace{0.07in}\alpha \rho \sigma }-\frac{1}{2}%
C_{\nu \alpha \rho \sigma }^{(2)}R_{\mu }^{\hspace{0.07in}\alpha \rho \sigma
}+\frac{1}{6}g_{\mu \nu }C_{3}\right.  \nonumber \\
&&\left. -\frac{1}{6}\nabla _{\mu }\nabla _{\nu }C_{2}+\frac{1}{6}g_{\mu \nu
}\nabla ^{2}C_{2}+\frac{1}{6}R_{\mu \nu }C_{2}\right) +\mathcal{O}(\nabla
^{4}C^{2})+\mathcal{O}(\nabla ^{2}C^{3})+\mathcal{O}(C^{4}),  \label{tgd4}
\end{eqnarray}
where 
\[
C_{\mu \nu \rho \sigma }^{(2)}=C_{\mu \nu \alpha \beta }C_{\hspace{0.12in}%
\rho \sigma }^{\alpha \beta },\qquad C_{2}=C_{\mu \nu \alpha \beta }C^{\mu
\nu \alpha \beta }. 
\]
In the list of higher orders that appears in the second line of (\ref{tgd4})
it is understood that pairs of covariant derivatives can be replaced by
curvature tensors, so $\mathcal{O}(\nabla ^{4}C^{2})=\mathcal{O}(\nabla
^{2}RC^{2})$, etc.

As promised, when the metric tensor is expanded around the metric $\overline{%
g}_{\mu \nu }$ of a space of constant curvature, an FLRW metric, or more
generally a conformally flat metric, then the quadratic part of the expanded
action $S_{\text{QG}}$ does not contain higher derivatives. We can prove
this fact considering the variation of $T_{\mu \nu }^{(g)}$ with respect to
the metric. The first two lines of (\ref{tg}) give contributions that
contain at most two derivatives of the fluctuation. The third line of (\ref
{tg}) gives contributions that are proportional to the Weyl tensor,
therefore vanish on conformally flat metrics. If $\overline{g}_{\mu \nu }$
does not belong to these classes of backgrounds then the quadratic part of
the action may contain higher derivatives. In general vertices do contain
higher derivatives of $g_{\mu \nu }$, multiplied by the couplings $\lambda
_{n}$.

To understant how to deal with such higher derivatives, recall that
renormalization, which is responsible for turning on the couplings $\lambda
_{n}$, is purely perturbative. To be consistent, the action $S_{\text{QG}}$
must be treated perturbatively in the $\lambda _{n}$s. In particular, we
must search for solutions of the field equations that are analytic in the $%
\lambda _{n}$s, at least away from singularities. Such solutions exist and
are uniquely determined by their limits $\lambda _{n}\rightarrow 0$. Indeed,
the field equations contain at most two time derivatives at $\lambda _{n}=0$%
. Therefore, working perturbatively in $\lambda _{n}$ we can convert every
terms that contain higher time derivatives into terms that contain at most
two time derivatives. In this way we obtain new field equations that are
perturbatively equivalent to (\ref{fieldequations}). Explicit examples of
this procedure are illustrated in section 4, when we study solutions of
black-hole type.

Similar methods are commonly used to eliminate runaway solutions caused by
higher-time derivatives, as in the case of the Abraham-Lorentz force in
classical electrodynamics \cite{jackson}. For applications to gravity see
refs. \cite{bel,parkersimon,acausal}. The elimination of unphysical
solutions has a price, because it generates violations of microcausality 
\cite{jackson}. We discuss these issues in detail at the end of section 4.

These facts, together with the presence of infinitely many independent
couplings, are there to remind us that $S_{\text{QG}}$ is not the action of
a fundamental theory, but must be viewed as an effective action that can be obtained from 
a more complete theory in a particular limit or 
integrating out some massive fields. In the same way as the Fermi
theory of weak interactions helped building the Standard Model, studying the
properties of $S_{\text{QG}}$ can be useful to identify the missing ultimate
theory of quantum gravity, which should be unitary, causal and
renormalizable with a finite number of independent couplings.
 
\section{Exact solutions of the field equations}

\setcounter{equation}{0}

In this section we study exact solutions of the $S_{\text{QG}}$-field
equations and relate them to known solutions of the Einstein field
equations. Because of the theorems proved in section 5 any solution of $S_{%
\text{QG}}$ can be perturbatively mapped into a solution of the field
equations of any action that is perturbatively equivalent to $S_{\text{QG}}$%
, for example $S_{\text{loc}}$ and $\tilde{S}_{\text{QG}}$.

We begin observing that in four dimensions all conformally flat metrics that
solve the Einstein equations 
\begin{equation}
R_{\mu \nu }-\frac{1}{2}Rg_{\mu \nu }-\Lambda g_{\mu \nu }=\kappa ^{2}T_{\mu
\nu },  \label{red}
\end{equation}
also solve the $S_{\text{QG}}$-field equations (\ref{fieldequations}), and
vice versa. The reason is that when $d=4$ and $C_{\mu \rho \nu \sigma }=0$
formulas (\ref{tg}) and (\ref{tgd4}), ensure that the gravitational
self-energy-momentum tensor $T_{\mu \nu }^{(g)}$ identically vanishes.
Moreover, the variation of $T_{\mu \nu }^{(g)}$ with respect to the metric
is proportional to the Weyl tensor, therefore it also vanishes on
conformally flat metrics. If we expand the metric tensor around conformally
flat backgrounds that solve (\ref{fieldequations}) in four dimensions the
propagator coincides with the one of Einstein gravity (if the same
gauge-fixing is used).

Now, if $d\Omega _{d-2}^{2}$ denotes the standard metric of the $(d-2)$-dimensional sphere, the metrics 
\begin{equation}
ds^{2}=g_{\mu \nu }dx^{\mu }dx^{\nu }=dt^{2}-a^{2}(t)\left( \frac{dr^{2}}{%
1-kr^{2}}+r^{2}d\Omega _{d-2}^{2}\right)  \label{flrwmetr}
\end{equation}
of homogeneous and isotropic spaces are conformally flat in arbitrary
dimensions. Indeed, it is easy to prove that the Weyl tensor vanishes
everywhere. The FLRW\ metrics have the form (\ref{flrwmetr}) and solve (\ref
{red}) with a homogeneous and isotropic distribution of matter, described by
an energy-momentum tensor $T_{\mu }^{\nu }$ equal to 
\begin{equation}
T_{\mu }^{\nu }(\rho ,p)=\rho \delta _{\mu }^{0}\delta _{0}^{\nu
}-p\sum_{i=1}^{d-1}\delta _{\mu }^{i}\delta _{i}^{\nu },  \label{trp}
\end{equation}
where the energy density $\rho $ and the pressure $p$ can be time-dependent.

Thus, the FLRW\ metrics are exact solutions of the $S_{\text{QG}}$-field
equations (\ref{fieldequations}) in four dimensions.

In higher dimensions we have to take the term $\int \sqrt{-g}\mathrm{\hat{G}}
$ into account. Nevertheless, in the classes of FLRW\ metrics and
conformally flat metrics we can find metric-independent maps that convert
solutions of the Einstein equations into solutions of the $S_{\text{QG}}$%
-field equations, and vice versa.

\paragraph{FLRW solutions in arbitrary dimensions\newline
}

Consider the $S_{\text{QG}}$-field equations (\ref{fieldequations}) with
matter energy-momentum tensor given by (\ref{trp}). We want to show that the
FLRW\ metrics (\ref{flrwmetr}) that solve 
\begin{equation}
R_{\mu \nu }-\frac{1}{2}Rg_{\mu \nu }-\Lambda g_{\mu \nu }=\kappa
^{d-2}T_{\mu \nu }(\tilde{\rho},\tilde{p})  \label{billy}
\end{equation}
also solve (\ref{fieldequations}), and vive versa, where $\tilde{\rho}$ and $%
\tilde{p}$ are suitable functions of $\rho $ and $p$. Inserting (\ref{billy}%
) into (\ref{fieldequations}) we find that this statement is true if and
only if 
\begin{equation}
T_{\mu \nu }(\tilde{\rho},\tilde{p})=T_{\mu \nu }(\rho ,p)+T_{\mu \nu
}^{(g)}.  \label{teq}
\end{equation}
Using (\ref{billy}) inside (\ref{tg}) (and recalling that $C_{\mu \nu \rho
\sigma }=0$) we easily get 
\[
T_{\hspace{0.12in}\mu }^{(g)\text{\hspace{0.01in}}\nu }=\Lambda _{0}\tilde{%
\rho}\left( \tilde{\rho}\delta _{\mu }^{0}\delta _{0}^{\nu }-(\tilde{\rho}+2%
\tilde{p})\sum_{i=1}^{d-1}\delta _{\mu }^{i}\delta _{i}^{\nu }\right) , 
\]
where 
\[
\Lambda _{0}=2\lambda _{0}\kappa ^{d}\frac{(d-3)(d-4)}{(d-2)(d-1)}, 
\]
therefore equation (\ref{teq}) is equivalent to the pair of
metric-independent quadratic equations 
\begin{equation}
\rho =\tilde{\rho}-\Lambda _{0}\tilde{\rho}^{2},\qquad p=\tilde{p}-\Lambda
_{0}\tilde{\rho}(\tilde{\rho}+2\tilde{p}),  \label{rpeq}
\end{equation}
for $\tilde{\rho}$ and $\tilde{p}$.

Given $\rho $ and $p$, we determine $\tilde{\rho}$ and $\tilde{p}$ solving
the equations (\ref{rpeq}). Then the usual FLRW solution with energy density 
$\tilde{\rho}$ and pressure $\tilde{p}$ solves the $S_{\text{QG}}$-field
equations with energy density $\rho $ and pressure $p$. Assuming $\rho
\Lambda _{0},p\Lambda _{0}\ll 1$ the solution can be worked out
perturbatively. For convenience, we report here the differential equations
satisfied by $a,\rho $ and $p$ in arbitrary dimensions: 
\[
\frac{\ddot{a}}{a}=\frac{2\Lambda -(d-1)\tilde{p}\kappa ^{d-2}-(d-3)\tilde{%
\rho}\kappa ^{d-2}}{(d-1)(d-2)},\qquad \frac{\mathrm{d}\tilde{\rho}}{\mathrm{%
d}t}=-(d-1)(\tilde{p}+\tilde{\rho})\left( \frac{\dot{a}}{a}\right) . 
\]
The cases $d=3,4$ can be seen as particular cases of the more general
solution.

Observe that in higher dimensions when we expand the metric around FLRW\
backgrounds the propagator does not coincide with the one obtained in
Einstein gravity (even if we use the same gauge-fixing). Nevertheless,
formula (\ref{tg}) shows that the quadratic part of the expanded action $S_{%
\text{QG}}$ does not contain higher derivatives. Indeed, it is just affected
by terms $\sim \tilde{\rho}\bigtriangledown ^{2}$ and $\sim \tilde{p}%
\bigtriangledown ^{2}$, and terms with fewer derivatives.

\paragraph{Conformally flat solutions in arbitrary dimensions\newline
}

More generally, if $\tilde{T}_{\mu }^{\nu }$ and $T_{\mu }^{\nu }$ are
related by the metric-independent polynomial equation 
\begin{equation}
T_{\mu }^{\nu }=\tilde{T}_{\mu }^{\nu }-\Lambda _{0}\frac{d-1}{d-2}\left( 2%
\tilde{T}_{\mu }^{\rho }\tilde{T}_{\rho }^{\nu }-\delta _{\mu }^{\nu }\tilde{%
T}_{2}-\frac{2}{d-1}\tilde{T}_{\mu }^{\nu }\tilde{T}+\frac{1}{d-1}\delta
_{\mu }^{\nu }\tilde{T}^{2}\right) .  \label{acco}
\end{equation}
where $\tilde{T}=\tilde{T}_{\rho }^{\rho }$ and $\tilde{T}_{2}=\tilde{T}%
_{\rho }^{\sigma }\tilde{T}_{\sigma }^{\rho }$, then the conformally flat
metrics that solve 
\begin{equation}
R_{\mu \nu }-\frac{1}{2}Rg_{\mu \nu }-\Lambda g_{\mu \nu }=\kappa ^{d-2}%
\tilde{T}_{\mu \nu }  \label{fis}
\end{equation}
also solve the $S_{\text{QG}}$-field equations, and vice versa. The
condition (\ref{acco}) is obtained inserting (\ref{fis}) into (\ref{tg}) and
(\ref{fieldequations}), and using $C_{\mu \nu \rho \sigma }=0$. Expanding
the metric tensor around a conformally flat solution the quadratic part of
the action $S_{\text{QG}}$ is free of higher derivatives.

\section{Approximate black-hole solutions}

\setcounter{equation}{0}

From the observational point of view, deformed black-hole solutions can
offer interesting possibilities to test modifications of general relativity.
Deviations from the Kerr metric, in particular, are the easiest to detect 
\cite{bambi}. Since black-hole solutions are not conformally flat, they are
affected in a non-trivial way by the corrections to Einstein gravity
contained in $S_{\text{QG}}$. In this section we study deformations of the
metrics of Schwarzschild and Kerr types.

We work in four dimensions and in the absence of matter, and keep only the
Goroff-Sagnotti constant $\Lambda _{\text{GS}}$, besides the Newton constant 
$G=\kappa ^{2}/8\pi $ and the cosmological constant $\Lambda $. The action
reads 
\begin{equation}
S_{\text{QG}}^{\prime }=-\frac{1}{2\kappa ^{2}}\int \sqrt{-g}\left(
R+2\Lambda +\frac{\Lambda _{\text{GS}}}{3}C_{3}\right) .  \label{spqg}
\end{equation}
We begin looking for spherically symmetric solutions of the form 
\begin{equation}
ds^{2}=\mathrm{e}^{\nu (r)+\omega (r)}dt^{2}-\mathrm{e}^{-\nu
(r)}dr^{2}-r^{2}(d\theta ^{2}+\sin ^{2}\theta \hspace{0.01in}d\varphi ^{2}).
\label{ansatz}
\end{equation}
It is worth mentioning that metrics of this type satisfy the peculiar
identity 
\begin{equation}
C_{\mu \nu \rho \sigma }^{(2)}=-\frac{\Omega }{2\sqrt{3}}C_{\mu \nu \rho
\sigma }+\frac{\Omega ^{2}}{12}(g_{\mu \rho }g_{\nu \sigma }-g_{\mu \sigma
}g_{\nu \rho }),\qquad \Omega ^{2}=C_{2},  \label{pecu}
\end{equation}
where the sign of $\Omega $ is determined to have $\Omega >0$ for the
Schwarzschild metric. This identity is useful to simplify various
expressions. Inserting the ansatz (\ref{ansatz}) into the field equations (%
\ref{fieldequations}) and using (\ref{tgd4}) we find differential equations
for $\nu (r)$ and $\omega (r)$. The $\Lambda _{\text{GS}}$-dependent
contributions involve up to four derivatives of these functions. Clearly,
higher-derivatives do not appear at $\Lambda _{\text{GS}}=0$ and, as
explained in section 2, we must search for solutions that are analytic in $%
\Lambda _{\text{GS}}$, at least away from singularities. Thus we can work
iteratively in $\Lambda _{\text{GS}}$, which allows us to convert the
higher-derivative terms into terms that have at most two derivatives. After
this conversion we find two (involved) equations of the form 
\begin{equation}
\nu ^{\prime }=F_{1}(\nu ,\omega ,r),\qquad \omega ^{\prime }=F_{2}(\nu
,\omega ,r),  \label{tru}
\end{equation}
for certain functions $F_{1}$ and $F_{2}$ that are analytic in $\Lambda _{%
\text{GS}}$, and two other equations that are automatically satisfied when (%
\ref{tru}) hold. We see that the solutions certainly exist and are uniquely
determined by their limits $\Lambda _{\text{GS}}\rightarrow 0$. However, we
do not have closed expressions for the functions $F_{1}$ and $F_{2}$,
therefore both the search for exact solutions and the numerical analysis
appear to be challenging tasks, also considering that the higher-derivative
form of the equations does not make numerical integration easy. Here we
content ourselves with the first perturbative corrections in $\Lambda _{%
\text{GS}}$.

Defining 
\[
\chi (r)=r\left( 1-\mathrm{e}^{\nu (r)}-\frac{\Lambda }{3}r^{2}\right) , 
\]
we find 
\[
\chi ^{\prime }=-\frac{2\Lambda _{\text{GS}}}{r^{7}}\chi ^{2}(16\chi
-15r+4\Lambda r^{3})+\mathcal{O}(\Lambda _{\text{GS}}^{2}),\qquad \omega
^{\prime }=\frac{24\Lambda _{\text{GS}}}{r^{7}}\chi ^{2}+\mathcal{O}(\Lambda
_{\text{GS}}^{2}). 
\]
To the lowest order of approximation, the solutions can be found replacing $%
\chi $ with a constant on the right-hand sides of these equations. We obtain 
\begin{eqnarray}
\mathrm{e}^{\nu (r)} &=&1-\frac{r_{s}}{r}-\frac{\Lambda }{3}r^{2}+\frac{%
6\Lambda _{\text{GS}}r_{s}^{2}}{r^{6}}\left( 1-\frac{8r_{s}}{9r}-\frac{4}{9}%
\Lambda r^{2}\right) +\mathcal{O}(\Lambda _{\text{GS}}^{2}),  \nonumber \\
\omega (r) &=&-\frac{4\Lambda _{\text{GS}}r_{s}^{2}}{r^{6}}+\mathcal{O}%
(\Lambda _{\text{GS}}^{2}),  \label{sc}
\end{eqnarray}
$r_{s}=2Gm$ being the usual Schwarzschild radius.

Using a computer program we worked out the metric up to the order $\Lambda _{%
\text{GS}}^{4}$ included. Higher-order corrections show that the solution
has the form 
\begin{equation}
-g_{rr}^{-1}=\mathrm{e}^{\nu (r)}=1-\frac{r_{s}}{r}-\frac{\Lambda }{3}r^{2}+%
\frac{r_{s}}{r}\sum_{n=1}^{\infty }\xi ^{n}P_{n},\qquad \omega (r)=\frac{%
r_{s}}{r}\sum_{n=1}^{\infty }\xi ^{n}Q_{n-1},  \label{formsol}
\end{equation}
where 
\[
\xi (r)=\frac{\Lambda _{\text{GS}}r_{s}}{r^{5}} 
\]
and $P_{n}$, $Q_{n}$ are polynomials of degree $n$ in $r_{s}/r$ and $\Lambda
r^{2}$. It is easy to verify that the expansion of $g_{tt}$ has the same
form as the one of $-g_{rr}^{-1}$. Thus the approximation obtained expanding
in powers of $\Lambda _{\text{GS}}$ is valid for $\xi \ll 1$, with $r_{s}/r$
and $\Lambda r^{2}$ bounded.

At $\Lambda =0$ the metric has an event horizon at a modified radius equal
to 
\begin{equation}
\bar{r}_{s}=r_{s}\left( 1-\frac{2}{3}\xi _{s}+\mathcal{O}(\xi
_{s}^{2})\right) ,  \label{radius}
\end{equation}
where $\xi _{s}=\xi (r_{s})$. The form (\ref{formsol}) of the solution shows
that both $g_{tt}$ and $g_{rr}^{-1}$ vanish at $r=\bar{r}_{s}$.

The informations we have gathered so far do not allow us to study the
curvature singularity at $r=0$. We just mention that once the action is
written in the form $S_{\text{QG}}$ it makes more sense to consider
curvature scalars such as $C_{2}$, $C_{3}$, etc., instead of the Kretschmann
scalar $R_{\mu \nu \rho \sigma }R^{\mu \nu \rho \sigma }$ (which coincides
with $C_{2}$ for Ricci flat metrics). Because of the identity (\ref{pecu})
we have 
\[
C_{3}=-\frac{1}{2\sqrt{3}}C_{2}^{3/2}. 
\]
We find (at $\Lambda =0$) 
\[
C_{2}=\frac{12r_{s}^{2}}{r^{6}}\left( 1-4\xi (r)\left( 12-13\frac{r_{s}}{r}%
\right) \right) +\mathcal{O}(\xi ^{2}). 
\]
To this order $C_{2}$ is equal to $R_{\mu \nu \rho \sigma }R^{\mu \nu \rho
\sigma }$, because the difference is quadratic in $R_{\mu \nu }$, therefore
at least $\mathcal{O}(\Lambda _{\text{GS}}^{2})$.

Now we switch to the modified Kerr metric. We study it at $\Lambda =0$ in
two limiting situations. We first consider slowly rotating black holes. To
the first order in $a=J/m$ at $\Lambda =0$, where $J$ is the angular
momentum, we find 
\[
ds^{2}=e^{\bar{\nu}(r)+\bar{\omega}(r)}dt^{2}-\mathrm{e}^{-\bar{\nu}%
(r)}dr^{2}-r^{2}(d\theta ^{2}+\sin ^{2}\theta d\phi ^{2})+2a\frac{r_{s}}{r}%
\left( 1+\frac{4\Lambda _{\text{GS}}r_{s}^{2}}{3r^{6}}\right) \sin
^{2}\theta dtd\phi , 
\]
plus $\mathcal{O}(\Lambda _{\text{GS}}^{2})$ and $\mathcal{O}(a^{2})$, where 
$\bar{\nu}$ and $\bar{\omega}$ are the same functions as before calculated
at $\Lambda =0$. The location of the event horizon is unmodified to this
order of approximation.

Moving one step forward, we study the large-distance expansion of the
deformed Kerr metric. Precisely, we take $r_{s},a\sim \varepsilon $ and $%
\Lambda _{\text{GS}}\sim \varepsilon ^{4}$, $\varepsilon \ll 1$ (i.e. we
assume that the constants $r_{s}$, $a$ and $\Lambda _{\text{GS}}$ are of
orders equal to their dimensions in units of coordinates), and calculate the
metric to the order $\varepsilon ^{8}$. Doing so, we automatically exclude
orders of $\Lambda _{\text{GS}}$ higher than the first. Indeed, $\Lambda _{%
\text{GS}}$ must always be multiplied by $r_{s}$, because $r_{s}=0$ gives
flat space. In Boyer-Lindquist coordinates we write 
\[
ds^{2}=g_{tt}dt^{2}+g_{rr}dr^{2}+g_{\theta \theta }d\theta ^{2}+g_{\phi \phi
}d\phi ^{2}+2g_{t\phi }dtd\phi 
\]
and obtain 
\begin{eqnarray}
g_{tt} &=&1-\frac{rr_{s}}{\rho ^{2}}+\frac{2\Lambda _{\text{GS}}r_{s}^{2}}{%
3\rho ^{8}}(3\rho ^{2}-2rr_{s}-54a^{2}\cos ^{2}\theta ),\qquad g_{\theta
\theta }=-\rho ^{2}+\frac{6a^{2}r_{s}^{2}}{\rho ^{6}}\Lambda _{\text{GS}%
}\sin ^{2}\theta ,  \nonumber \\
g_{rr} &=&-\frac{\rho ^{2}}{\Delta }+\frac{2\Lambda _{\text{GS}}r_{s}^{2}}{%
3\rho ^{6}\Delta }\left( 9a^{2}+9\rho ^{2}+rr_{s}+r_{s}^{2}-297a^{2}\cos
^{2}\theta \right) ,  \label{modK} \\
g_{t\phi } &=&\frac{arr_{s}}{\rho ^{2}}\sin ^{2}\theta \left( 1+\frac{%
4\Lambda _{\text{GS}}r_{s}^{2}}{3\rho ^{6}}\right) ,\qquad g_{\phi \phi
}=-\sin ^{2}\theta \left( a^{2}+r^{2}+\frac{a^{2}rr_{s}}{\rho ^{2}}\sin
^{2}\theta \right) ,  \nonumber
\end{eqnarray}
where, as usual, 
\[
\rho ^{2}=r^{2}+a^{2}\cos ^{2}\theta ,\qquad \Delta =r^{2}-rr_{s}+a^{2}. 
\]

Observe that the modified Kerr metric (\ref{modK}) is more general than the
deformed metrics considered in ref. \cite{psaltis}, where deviations from
Kerr are parametrized by one function $h$ of $r$ and $\theta $. Because of
this, calculations are rather involved. Using a computer program, five
independent functions of $r$ and $\theta $ have been used to work out the
approximate solution given above. Note that at the end there is no
deformation of $g_{\phi \phi }$.

We stress again that renormalization predicts $\Lambda _{\text{GS}}\neq 0$,
therefore the deviations worked out in this section can be viewed as
predictions of quantum gravity. Their practical detectability depends on the
actual value of the constant $\Lambda _{\text{GS}}$. Theoretically, we
cannot predict the value of $\Lambda _{\text{GS}}$, but only the $\Lambda _{%
\text{GS}}$-running, which gives us an estimate of the minimum value of $%
|\Lambda _{\text{GS}}|$. Using the two-loop result of \cite{sagnotti} we
find 
\[
\Delta \Lambda _{\text{GS}}(\ell ,\ell ^{\prime })=\Lambda _{\text{GS}}(\ell
)-\Lambda _{\text{GS}}(\ell ^{\prime })=\frac{209l_{P}^{4}}{30(4\pi )^{2}}%
\ln \frac{\ell }{\ell ^{\prime }}, 
\]
where $l_{P}=\sqrt{G}$ is the Planck length and $\Lambda _{\text{GS}}(x)$ is
the running coupling at the scale $x$. If we take $\ell $ equal to the
diameter of the observable universe and $\ell ^{\prime }$ equal to the
Planck length itself, we obtain 
\[
|\Delta \Lambda _{\text{GS}}|\sim 6l_{P}^{4}. 
\]
If the value of $|\Lambda _{\text{GS}}|$ were around $6l_{P}^{4}$ there
would be no chance to detect the deviations we have worked out so far. We
can only hope that $|\Lambda _{\text{GS}}|$ has a much larger value in
nature. Light black holes are the ones that are affected more sensibly.
Taking a mass equal to 5 solar masses, we need at least 
\begin{equation}
|\Lambda _{\text{GS}}|\sim 10^{156}l_{P}^{4}=10^{44}(\mathrm{eV})^{-4}
\label{theo}
\end{equation}
to get $\xi _{s}\sim 1$. In this case the deviations would be appreciable
right outside the black hole. The Schwarschild radius (\ref{radius}) would
be modified in a sensible way and effects on the deflection of light, for
example, could be detected. Depending on the precision of our instruments,
smaller values of $\xi _{s}$ could suffice. In case no deviations are
observed it is possible to put experimental bounds on $\Lambda _{\text{GS}}$%
. Observe that as long as $|\Lambda _{\text{GS}}|$ is much larger than $%
6l_{P}^{4}$, for all practical purposes $\Lambda _{\text{GS}}$ does not run
throughout the universe.

So far we have studied static and stationary solutions, but if we are
interested in metrics that depend on time, as well as the motion of light
and particles in the metrics we have found, we must discuss the violations
of causality induced by the presence of higher time derivatives.

To understand the problem it is useful to briefly recall the case of the
Abraham-Lorentz force \cite{jackson} in classical electrodynamics, where the
radiation emitted by an accelerated charged particle of mass $m$ is
described by one of the equations 
\begin{equation}
m\left( 1-\tau \frac{\mathrm{d}}{\mathrm{d}t}\right) a(t)=F(t),\qquad
ma(t)=\langle F(t)\rangle \equiv \frac{1}{\tau }\int_{t}^{\infty }\mathrm{d}%
t^{\prime }\mathrm{e}^{(t-t^{\prime })/\tau }F(t^{\prime }),  \label{AL}
\end{equation}
where $\tau =2e^{2}/(3mc^{3})$, $a$ is the acceleration and $F$ is an
external force. The first equation is the standard, higher-derivative one.
The second equation is obtained from the first one with the same procedure
used to obtain (\ref{tru}), i.e. demanding analyticity in $\tau $. This
requirement eliminates the runaway solution, but generates a violation of 
\textit{micro}causality. Indeed, to determine the motion at a given time $t$
we must know the external force at future times $t^{\prime }$ such that $%
t\leqslant t^{\prime }\lesssim t+\tau $. On the other hand, if $F(t^{\prime
})\neq 0$ only for $0\leqslant t^{\prime }\leqslant T$ all events appear to
be causal at any time $t>T$.

Let us now turn to the case of gravity. Even if the metric deviations
predicted here were detected, they would not necessarily provide an indirect
evidence that microcausality is violated. The reason is that the action $S_{%
\text{QG}}$ is most probably the effective theory of a more complete, causal
theory. It could be obtained, for example, integrating out some degrees of
freedom. That said, to detect violations of microcausality we should catch
acausal events in the act, compare a sufficient number of different
situations, and prove that no causal equations can explain the data.

Considering a fluctuation $\delta g$ around the metric given by (\ref{ansatz}%
) and (\ref{sc}), higher-time derivative terms provided by $\delta T_{\mu
\nu }^{(g)}$ are mutiplied by the Weyl tensor $C\sim r_{s}/r^{3}$ or by $%
\nabla C\sim r_{s}/r^{4}$: 
\[
\kappa ^{2}\delta T_{\mu \nu }^{(g)}\sim \Lambda _{\text{GS}}\nabla C\nabla
^{3}\delta g+\Lambda _{\text{GS}}C\nabla ^{4}\delta g 
\]
Comparing these terms with the ones contained in the Einstein field
equations and assuming that the derivatives of $\delta g$ are time ones, for 
$\xi (r)<1$ causality violations last for a typical time equal to 
\[
\tau (r)=r\sqrt{\xi (r)}. 
\]
In the case of gravitational lensing by a light black hole, taking $r$
around a few times $r_{s}$ and assuming $\xi (r)\sim 1$ it is necessary to
resolve time intervals of about 10$^{-4}$ seconds.

\section{Perturbative equivalence of actions and solutions of the field
equations}

\setcounter{equation}{0}

Renormalization cannot determine the action unambiguously. It only
determines the \textit{perturbative equivalence} class to which the action
belongs. We say that two actions $S_{1}$ and $S_{2}$ are perturbatively
equivalent if

$i$) \textit{they are perturbative expansions around the same unperturbed
action }$\bar{S}$\textit{\ and }

$ii$) \textit{they can be mapped into each other by means of perturbative
field redefinitions and parameter redefinitions.}

A perturbative field redefinition is a field redefinition that can be
expressed as the identity map plus a perturbative series of local monomials
of the fields and their derivatives. Using an appropriate field-covariant
formalism \cite{fieldcov} perturbative field redefinitions can be
implemented in functional integrals as true changes of integration
variables, instead of mere replacements of integrands. Generating
functionals, suitably generalized \cite{masterf,mastercan}, behave as
scalars.

The actions $S_{\text{QG}}$, $\tilde{S}_{\text{QG}}$ and $S_{\text{loc}}$
are perturbatively equivalent. They are mapped into one another by
perturbative redefinitions of the metric tensor and redefinitions of the
parameters $\lambda $ and $\zeta $, where $\zeta $ denote the parameters of
the matter action $S_{m}$. As a consequence, the solutions of their field
equations can also be perturbatively mapped into one another. In this
section we study the map in detail.

We begin with the perturbative equivalence of $S_{\text{QG}}$ and $\tilde{S}%
_{\text{QG}}$. There exists a redefinition of the metric tensor of the form 
\begin{equation}
g=g^{\prime }+\mathcal{O}(\hat{R}^{2}),  \label{fref}
\end{equation}
where $\hat{R}$ denotes the tensor (\ref{hatc}) and its contractions, and
parameter redefinitions $\lambda ^{\prime }$,$\zeta ^{\prime }$ such that 
\begin{equation}
S_{\text{QG}}(g,\varphi ,\Lambda ,\lambda ,\zeta )=\tilde{S}_{\text{QG}%
}(g^{\prime },\varphi ,\Lambda ,\lambda ^{\prime },\zeta ^{\prime }).
\label{tesi}
\end{equation}

We work inductively in the power $n_{R}$ of Weyl or hatted curvature
tensors. Specifically, we assume 
\begin{equation}
S_{\text{QG}}(g,\varphi ,\Lambda ,\lambda ,\zeta )=\tilde{S}_{\text{QG}%
}(g^{\prime },\varphi ,\Lambda ,\lambda ^{\prime },\zeta ^{\prime })+Y_{\bar{%
n}_{R}+1},  \label{ind}
\end{equation}
where $g$ and $g^{\prime }$ are related by a field redefinition of the form (%
\ref{fref}), $\bar{n}_{R}\geqslant 2$ and $Y_{\bar{n}_{R}+1}$ is
matter-independent and $\mathcal{O}(\hat{R}^{\bar{n}_{R}+1})$. The identity (%
\ref{ind}) is obviously satisfied for $\bar{n}_{R}=2$. It is sufficient to
show that formula (\ref{ind}) with arbitrary $\bar{n}_{R}\geqslant 2$
implies a similar relation with $\bar{n}_{R}\rightarrow \bar{n}_{R}+1$.

Consider the terms of $Y_{\bar{n}_{R}+1}$ that have precisely $\bar{n}_{R}+1$
hatted curvature tensors. Express $\hat{R}_{\mu \nu \rho \sigma }$ in terms
of the Weyl tensor, $\hat{R}_{\mu \nu }$ and $\hat{R}$. The terms containing
only Weyl tensors can be mapped defining relations between the appropriate
parameters $\lambda $ and $\lambda ^{\prime }$. Once we have done this, we
obtain 
\[
S_{\text{QG}}(g,\varphi ,\Lambda ,\lambda ,\zeta )=\tilde{S}_{\text{QG}%
}(g^{\prime },\varphi ,\Lambda ,\lambda ^{\prime \prime },\zeta ^{\prime })+%
\tilde{Y}_{\bar{n}_{R}+1}, 
\]
where $\tilde{Y}_{\bar{n}_{R}+1}=\mathcal{O}(\hat{R}^{\bar{n}_{R}+1})$ is
still matter-independent, but now it is also proportional to $\hat{R}_{\mu
\nu }$ or $\hat{R}$. We can write 
\[
\tilde{Y}_{\bar{n}_{R}+1}=\int \left( \hat{R}_{\mu \nu }-\frac{1}{2}g_{\mu
\nu }\hat{R}\right) X_{\bar{n}_{R}}^{\mu \nu },\qquad X_{\bar{n}_{R}}^{\mu
\nu }=\mathcal{O}(\hat{R}^{\bar{n}_{R}}). 
\]
Using (\ref{fieldequations}) and (\ref{tg}), adapted to $\tilde{S}_{\text{QG}%
}$, the variation $\tilde{E}_{\text{QG}}^{\mu \nu }$ of $\tilde{S}_{\text{QG}%
}$ with respect to the metric tensor can be written in the form 
\[
\tilde{E}_{\text{QG}}^{\mu \nu }=\hat{R}^{\mu \nu }-\frac{1}{2}g^{\mu \nu }%
\hat{R}+E_{m}^{\mu \nu }+\mathcal{O}(\hat{R}^{2}), 
\]
where $E_{m}^{\mu \nu }$ is the analogous variation of $S_{m}$, therefore 
\[
\tilde{Y}_{\bar{n}_{R}+1}=\int \tilde{E}_{\text{QG}}X_{\bar{n}_{R}}+Y_{m,%
\bar{n}_{R}}+\mathcal{O}(\hat{R}^{\bar{n}_{R}+2}), 
\]
$Y_{m,\bar{n}_{R}}$ denoting $\mathcal{O}(\hat{R}^{\bar{n}_{R}})$-terms
proportional to the matter fields $\varphi $. At this point, we have 
\[
S_{\text{QG}}(g,\varphi ,\Lambda ,\lambda ,\zeta )=\tilde{S}_{\text{QG}%
}(g^{\prime },\varphi ,\Lambda ,\lambda ^{\prime \prime },\zeta ^{\prime
})+\int \tilde{E}_{\text{QG}}^{\prime }X_{\bar{n}_{R}}^{\prime }+Y_{m,\bar{n}%
_{R}}^{\prime }+\mathcal{O}(\hat{R}^{\bar{n}_{R}+2}), 
\]
where $\tilde{E}_{\text{QG}}^{\prime }$, $X_{\bar{n}_{R}}^{\prime }$ and $%
Y_{m,\bar{n}_{R}}^{\prime }$ are $\tilde{E}_{\text{QG}}$, $X_{\bar{n}_{R}}$
and $Y_{m,\bar{n}_{R}}$ once the metric tensor $g$ is expressed in terms of $%
g^{\prime }$. Consider the redefinition 
\[
g^{\prime \prime }=g^{\prime }+X_{\bar{n}_{R}}^{\prime } 
\]
of the metric tensor. We have 
\[
S_{\text{QG}}(g,\varphi ,\Lambda ,\lambda ,\zeta )=\tilde{S}_{\text{QG}%
}(g^{\prime \prime },\varphi ,\Lambda ,\lambda ^{\prime \prime },\zeta
^{\prime })+Y_{m,\bar{n}_{R}}^{\prime \prime }+\mathcal{O}(\hat{R}^{\bar{n}%
_{R}+2}), 
\]
where we have used $\bar{n}_{R}\geqslant 2$. Finally, the terms $Y_{m,\bar{n}%
_{R}}$ can be reabsorbed redefining the parameters $\zeta ^{\prime }$.
Therefore there exist $\zeta ^{\prime \prime }$ such that 
\[
S_{\text{QG}}(g,\varphi ,\Lambda ,\lambda ,\zeta )=\tilde{S}_{\text{QG}%
}(g^{\prime \prime },\varphi ,\Lambda ,\lambda ^{\prime \prime },\zeta
^{\prime \prime })+\mathcal{O}(\hat{R}^{\bar{n}_{R}+2}). 
\]
This relation promotes the inductive hypothesis (\ref{ind}) from $\bar{n}%
_{R} $ to $\bar{n}_{R}+1$, which proves the theorem.

The same procedure can be used to modify the $\mathcal{O}(\hat{R}^{3})$%
-sector of the action $\tilde{S}_{\text{QG}}$ adding any $\mathcal{O}(\hat{R}%
^{3})$-terms proportional to the hatted Ricci tensor.

Now we show the perturbative equivalence of $S_{\text{QG}}$ and $S_{\text{loc%
}}$. These actions differ by terms quadratically proportional to the hatted
Ricci tensor and $\mathcal{O}(\hat{R}^{3})$-terms proportional to the hatted
Ricci tensor. To quickly prove their equivalence we use a theorem derived in
ref. \cite{acausal}, stating that any terms quadratically proportional to
the field equations can be reabsorbed into a perturbative field
redefinition. In particular, there exists a field redefinition 
\begin{equation}
\tilde{g}=g+\mathcal{O}(\hat{R}_{\mu \nu })  \label{map}
\end{equation}
such that 
\[
-\frac{1}{2\kappa ^{d-2}}\int \sqrt{-g}(R+2\Lambda )+\frac{1}{2}\int \sqrt{-g%
}\hspace{0.01in}\hat{R}_{\mu \nu }Q^{\mu \nu \rho \sigma }\hat{R}_{\rho
\sigma }=-\frac{1}{2\kappa ^{d-2}}\int \sqrt{-\tilde{g}}\left( R(\tilde{g}%
)+2\Lambda \right) , 
\]
where $Q$ is any perturbatively local derivative operator. Using this map
and the properties of $\int \sqrt{-g}\mathrm{\hat{G}}$, in particular its
variation with respect to the metric, encoded in (\ref{tg}), we can convert $%
S_{\text{loc}}$ into an action $\tilde{S}_{\text{QG}}$ with unrestricted
scalars $\mathcal{I}_{n}^{(\Lambda )}$. Then using the map (\ref{fref}) and
parameter-redefinitions we can convert the action to $S_{\text{QG}}$.

Finally, recall that when maps such as (\ref{fref}) and (\ref{map}) lower
the number of time derivatives, they also generate violations of
microcausality \cite{acausal,halat}, to which the arguments of the previous
section apply.

\section{Conclusions}

\setcounter{equation}{0}

The action of quantum gravity is determined by renormalization. It can be
simplified dropping terms proportional to the hatted Ricci tensor, because
those terms can be reabsorbed into perturbative field redefinitions and
parameter redefinitions. Doing so, we can arrange the action in different
perturbatively equivalent ways, which may help us uncover different
properties, identify different classes of exact solutions, or reduce the
effort to study approximate solutions. We singled out a convenient form $S_{%
\text{QG}}$ that allows us, to some extent, to have control on the
infinitely many couplings of the theory. Among the other things, we can show
that some well known metrics, such as the FLRW metrics, are exact solutions
of the field equations or can be mapped into exact solutions. Precisely, in
four dimensions the solutions coincide with the usual ones, while in
dimensions greater than four they coincide with the usual ones once the
density and the pressure are mapped into simple functions of themselves.
More generally, all conformally flat solutions of Einstein gravity can be
mapped in a metric-independent way into conformally flat solutions of $S_{%
\text{QG}}$, and vice versa. The quadratic terms of the action, generated
expanding the metric around these solutions, are free of higher derivatives.
Solutions that are not conformally flat are instead modified in a nontrivial
way. We have studied the first corrections to the metrics of Schwarzschild
and Kerr types, expanding in powers of the Goroff-Sagnotti constant.

Vertices can contain arbitrarily high derivatives of the metric tensor. The
solutions of the field equations that are analytic in the couplings $\lambda
_{n}$, at least away from singularities, are uniquely determined by initial conditions
of Cauchy type. However, those solutions violate microcausality. These features and the presence
of infinitely many independent couplings point towards a missing, more
fundamental theory, which should be unitary, causal and renormalizable with
a finite number of independent couplings.

Most of the properties we have studied originate from high-energy physics,
specifically renormalization. However, they may have effects detectable in
astrophysical observations. For example, it would be desirable to compare
predictions and observational data to put constraints on the magnitude of
the Goroff-Sagnotti constant. Renormalization only tells us that this
constant is non-vanishing. A further reason to motivate investigations 
of the properties of $S_{\text{QG}}$ is that they could
help us identify the ultimate theory of quantum gravity, in the same way as
the Fermi theory of weak interactions was helpful to build the Standard
Model.

\vskip .6truecm \textbf{Acknowledgments}

\vskip .3truecm

I thank C. Bambi, G. Cella, P. Menotti and L. Modesto for inspiring
discussions and correspondence. I also thank the Physics Department of Fudan
University, Shanghai, for hospitality during the initial stage of this work.

\setcounter{section}{0} \renewcommand{\thesection}{\Alph{section}}

\section{Appendix: Equivalent truncations of the action}

\setcounter{equation}{0}\renewcommand{\theequation}{\thesection.%
\arabic{equation}}

Since the number of parameters is infinite, it is useful to define
appropriate truncations to classify the invariants and expand
perturbatively, consistently with the diagrammatic expansion.

We study Feynman diagrams expanding the metric as 
\begin{equation}
g_{\mu \nu }=\bar{g}_{\mu \nu }+\kappa ^{(d-2)/2}h_{\mu \nu },  \label{espa}
\end{equation}
where the background $\bar{g}_{\mu \nu }$ is a conformally flat solution of
the field equations, in the case of $S_{\text{QG}}$, or a space of constant
curvature, in the cases of $\tilde{S}_{\text{QG}}$ and $S_{\text{loc}}$.
Invariance of the functional integral under translations ensures that the
results we obtain do not depend on the choice of background $\bar{g}_{\mu
\nu }$. The graviton propagator is determined by the Hilbert term and the
cosmological term. It depends on $\bar{g}_{\mu \nu }$ and $\Lambda $, but
not on $\kappa $ and the $\lambda _{n}$s. The propagators of matter fields
can of course depend on masses $m$. Let $E$ denote the overall energy scale
of correlation functions. We assume $\kappa E,\kappa m,\kappa |\Lambda
|^{1/2}\ll 1$, and that the values of $\lambda _{n}$ are bounded from above
(namely there exists a constant $M$ such that $|\lambda _{n}|<M$ for every $%
n $). We do not assume particular inequalities among $E$, $m$ and $\Lambda $%
, in the same way as we normally do not expand Feynman diagrams in powers of 
$m $ or $1/m$. Thus for our purposes $E$, $m$ and $\Lambda $ can be assumed
to be of the same order. Vertices are multiplied by powers of $\kappa $, $%
\Lambda $ and $\lambda _{n}$. Apart from this kind of factors, Feynman
diagrams give integrals that depend only on $\bar{g}_{\mu \nu }$, $m$ and $%
\Lambda $. Therefore, by the Wick theorem and power-counting we can write $%
h_{\mu \nu }\sim E$ (or $h_{\mu \nu }\sim |\Lambda |^{1/2},m$).

The diagrammatic expansion is an expansion in powers of $\kappa E$, $\kappa
|\Lambda |^{1/2}$ and $\kappa m$. 
A truncation of the diagrammatic expansion amounts to
discard powers of these quantities larger than some $T$. Below we concentrate on the gravitational sector, since the matter sector can be treated similarly. Moreover, we identify $\Lambda$ and $m^2$. Precisely, we
classify the contributions as 
\begin{equation}
\kappa ^{-d}(\kappa ^{2}\Lambda )^{a}(\kappa \bar{\nabla})^{b}(\kappa
^{(d-2)/2}h)^{c},  \label{espar}
\end{equation}
where $\bar{\nabla}$ denotes the covariant derivative in the background $%
\bar{g}_{\mu \nu }$, and pairs of $\bar{\nabla}$s can also stand for
curvature tensors $\bar{R}$. The number $c$ is the number of external legs
of the diagram (or vertex, at the tree level), while $b$ is the power of
(external) momenta and $a$ is the power of $\Lambda $. Higher powers of $%
\kappa ^{2}\Lambda $ can be generated by radiative corrections and
renormalize the parameters $\lambda _{n}$. The truncation to order $T$ is
obtained discarding the contributions that have 
\begin{equation}
2a+b+\frac{d-2}{2}c>T.  \label{c1}
\end{equation}
This kind of truncation preserves general covariance only within the
truncation, namely up to powers $T^{\prime }>T$ of $\kappa E$, $\kappa
|\Lambda |^{1/2}$ and $\kappa m$. Clearly, the Feynman diagrams that
contribute within the truncation are constructed with a finite number of
vertices. Moreover, they are themselves finitely many, since every loop
raises the power of $\kappa $. We call this truncation \textit{diagrammatic
truncation}.

There is actually an alternative truncation \cite{abse}, which simply
amounts to truncate the sums appearing in (\ref{cqgac}), (\ref{sloc}) and (%
\ref{oldac}) to finite numbers of terms. Its advantage is that it is
manifestly general covariant, although its connection with Feynman diagrams
is less apparent. Precisely, we discard, according to the case ($S_{\text{QG}%
}$ or $\tilde{S}_{\text{QG}}$-$S_{\text{loc}}$), the terms 
\begin{equation}
\sim \kappa ^{-d}(\kappa ^{2}\Lambda )^{n_{\Lambda }}(\kappa \nabla
)^{n_{\nabla }}(\kappa ^{2}C)^{n_{R}},\qquad \text{or}\qquad \sim \kappa
^{-d}(\kappa ^{2}\Lambda )^{n_{\Lambda }}(\kappa \nabla )^{n_{\nabla
}}(\kappa ^{2}\hat{R})^{n_{R}},  \label{struc}
\end{equation}
with 
\begin{equation}
2n_{\Lambda }+n_{\nabla }+2n_{R}>N,  \label{c2}
\end{equation}
for some $N$. Expanding the structures (\ref{struc}) according to (\ref{espa}%
) we get terms (\ref{espar}) with 
\[
a=n_{\Lambda },\qquad b=n_{\nabla }+2n_{R},\qquad c=n_{R}+q, 
\]
where $q\geqslant 0$ is integer. We can choose a basis such that each
invariant $\int \sqrt{-g}\hspace{0.01in}\mathcal{I}_{n}$, $\int \sqrt{-g}%
\hspace{0.01in}\mathcal{\bar{I}}_{n}^{(\Lambda )}$ and $\int \sqrt{-g}%
\hspace{0.01in}\mathcal{I}_{n}^{(\Lambda )}$ is uniquely determined by its $%
q=0$-contribution. The other contributions are then fixed by general
covariance.

The two truncations are actually equivalent, in the sense that a
diagrammatic truncation covers a certain general covariant truncation, and
vice versa. Let us describe how to switch back and forth between the two.
Since (\ref{c2}) implies (\ref{c1}) with $T=N$, the general covariant
truncation to order $N$ covers the diagrammatic truncation to order $N$. To
study the converse implication, we recall that, by general covariance, it is
enough to determine the $q=0$-contribution to an invariant to determine the
full invariant. Consider the terms (\ref{espar}) and analyze them for
increasing number of external legs $c$. Doing so, $q=0$-contributions coming
from new invariants can be disentangled from $q>0$-contributions coming from
invariants determined for smaller values of $c$. This procedure allows us to
determine the structures (\ref{struc}) with 
\[
n_{\Lambda }=a,\qquad n_{\nabla }=b-2c,\qquad n_{R}=c. 
\]
Because of (\ref{c1}), the terms we cannot determine satisfy 
\begin{equation}
2n_{\Lambda }+n_{\nabla }+2n_{R}>\frac{4}{d+2}\left( 2n_{\Lambda }+n_{\nabla
}+\frac{d+2}{2}n_{R}\right) >\frac{4T}{d+2}.  \label{gt}
\end{equation}
We conclude that the diagrammatic truncation to order $T$ covers the general
covariant truncation to order 
\[
N=\frac{4T}{d+2}. 
\]

\end{document}